\documentclass[12pt]{amsart}
\usepackage{xypic}
\usepackage{amsmath}
\usepackage{amssymb}
\usepackage{enumerate}
\usepackage{color}
\usepackage{pgf}
\usepackage{tikz}
\usetikzlibrary{arrows,automata}
\usepackage{relsize}
\usepackage{latexsym}
\usepackage{pgfplots}
\usetikzlibrary{positioning}
\usepackage[T1]{fontenc}
\usepackage{soul} 
\usepackage{amsfonts}
\usepackage[T1]{fontenc}
\usepackage{geometry}
\usepackage{amsmath}
\usepackage{amsthm}
\usepackage{stmaryrd}
\usepackage{stackrel}
\usepackage[all]{xy}
\theoremstyle{remark}

\makeatletter
\theoremstyle{plain}
\newtheorem{thm}{\protect\theoremname}[section]
\theoremstyle{definition}
\newtheorem{example}[thm]{\protect\examplename}
\theoremstyle{definition}
\newtheorem{defn}[thm]{\protect\definitionname}
\theoremstyle{plain}

\ifx\proof\undefined
\newenvironment{proof}[1][\protect\proofname]{\par
	\normalfont\topsep6\p@\@plus6\p@\relax
	\trivlist
	\itemindent\parindent
	\item[\hskip\labelsep\scshape #1]\ignorespaces
}{%
	\endtrivlist\@endpefalse
}
\providecommand{\proofname}{Proof}
\fi
\theoremstyle{plain}
\newtheorem{prop}[thm]{\protect\propositionname}
\theoremstyle{remark}

\date{}

\makeatother

\usepackage{babel}
\providecommand{\definitionname}{Definition}
\providecommand{\examplename}{Example}
\providecommand{\lemmaname}{Lemma}
\providecommand{\propositionname}{Proposition}
\providecommand{\remarkname}{Remark}
\providecommand{\theoremname}{Theorem}

\begin{document}
\title{Structural constraints to compare phenomenal experience}

\author[J. D\'iaz-Boils]{J. D\'iaz-Boils}
\address{Departament d'Economia Aplicada\\ Facultat d'Economia\\ Avinguda dels Tarongers\\ Universitat de Val\`encia\\ 46022-Val\`encia. Spain.}
\email{joaquin.diaz@uv.es}

\author[N. Tsuchiya]{N. Tsuchiya}
\address{School of Psychological Sciences, Monash University, Melbourne, Victoria, Australia\\ Turner Institute for Brain and Mental Health \& School of Psychological Sciences, Faculty of Medicine, Nursing, and Health Sciences, Monash University, Melbourne, Victoria, Australia\\ Center for Information and Neural Networks (CiNet), National Institute of Information and Communications Technology (NICT), Suita-shi, Osaka 565-0871, Japan\\ Laboratory Head Laboratory of Qualia Structure, ATR Computational Neuroscience Laboratories, 2-2-2 Hikaridai, Seika-cho, Soraku-gun, Kyoto 619-0288, Japan.}
\email{naotsugu.tsuchiya@monash.edu }

\author[CM. Signorelli]{CM. Signorelli}

\address{Department of Computer Science, University of Oxford, 15 Parks Rd, Oxford, OX1 3QD, United Kingdom\\ Center for Philosophy of Artificial Intelligence, University of Copenhagen, Karen Blixens Plads 8, Copenhagen, 2300, Denmark\\ Laboratory of Neurophysiology and Movement Biomechanics (LNMB), Universit\'e Libre de Bruxelles (ULB), Route de Lennik 808, CP 640. Building N, campus Erasme, 1070, Brussels, Belgium}

\email{cam.signorelli@cs.ox.ac.uk}



\maketitle

\begin{abstract}
The article introduces a partial order structure to study the relations between levels and contents of conscious experience within a formal and mathematical model. Inspired by Husserl's phenomenology, our account models structural aspects of experiences as layers that can be experienced in potentiality or in actuality, composing further experiences and imposing an order among them. From our simple set of assumptions, yet formal analysis, we conclude that assuming phenomenal comparison among experiences yields only a partial comparison. Some experiences are comparable (e.g. experiences that carry different aspects in actuality and potentiality), but others are not (e.g. experiences that carry the same number, but different aspects, in actuality). These formal results imply the evolution of consciousness as diverse modes of experiencing, instead of an absolute scale along species. Moreover, our methodology also elucidates structural constraints from conceptual assumptions to mathematical structure, and mathematical structure to interpretations about phenomenal experiences. This makes the mathematical formalisation transparent, opening new avenues to explore the consequences of formal models based on our partial order layer structure.

\\
\end{abstract}

\section{Introduction}\label{intro}

Applications of pure mathematics within consciousness science are leading us to new insights. Different mathematical structures, i.e. a set of operations and relations satisfying specific requirements or axioms, are applied to model, and formalise experiential structures, i.e. a set of relationships characterising either the interactions among subjective experiences or aspects of them. According to Husserl's phenomenology \cite{Husserl1969,Husserl2001}, after applying the phenomenological method, one may end up with phenomenological invariants, structural relations common to all conscious experiences \cite{Husserl1969,Husserl2001}. Structural accounts include analyses such as qualitative similarity (e.g. asking if X is closer to Y than Z, being X orange, Y red, and Z green, or any other representation) or temporal relations (e.g. experience X preceding another experience Z, see \cite{Husserl2001} for specific structures on time-consciousness), among others. In both cases, structural analyses and their eventual mathematisation can lead to general structures. We are interested in such structures (e.g., 'bigger than', 'closer', 'first than', 'next'), rather than specific experiential structures such as colour, taste, perception, or imagination (ongoing work instantiates the model in specific cases). In other words, we expect to mathematise aspects of experience in the most general case and focus on how these aspects relate and/or obey general rules of composition that can be instantiated in specific cases. Our methodology is therefore formal, meaning we can organise a phenomenon (or subclasses of it) under certain principles and structures to study the consequences of using established manipulation rules (e.g. definitions, theorems, postulates)\cite{Chartrand2018,Solow2014}.

Examples of this approach range from the use of category theory to unravel the distinctions between levels and content of consciousness \cite{Tsuchiya2021}, the use of monoidal categories to compose primitive aspects of experience \cite{Signorelli2020c,Signorelli2020f}, and the use of multi-layers to account for embodied experience \cite{Signorelli2024e,Sign}. Many other examples exist in the literature \cite{Prentner2019,Prakash2020,Tull2020,Phillips2020,Lee2021,Yoshimi2016,Mason2021,Resende2025,prentner2024category, doi:10.1073/pnas.2115934119}, each one using their preferred, yet convenient, mathematical structure \cite{Kleiner2024} (see \cite{Wachtel2022} for a criticism and limitations of this approach). 

In the following, we present a simplified version of the multi-layer model introduced in \cite{Signorelli2024e,Sign}, to show how conceptual assumptions constrain mathematical formalisations and, in turn, how these formalisations constrain our interpretations regarding comparisons of phenomenal experience. These constraints will be called \textit{structural constraints}. We will start by introducing the minimal elements for a mathematical structure of conscious experience in the form of multi-layers, and the distinction between phenomenological structures working as labels or constraints of physical ones (following insight from \cite{Husserl1969,Husserl2001}) (section \ref{multi}). This already reveals the power of a multi-layer model against other mathematical structures. Then, we will formalise cases of \textit{experiential comparison} (i.e. comparing structural aspects among experiences) by combining the assumptions behind the notions of levels of consciousness and content of consciousness in one single mathematical structure (section \ref{pomonoid}). This single structure enables us to investigate relational aspects of animal consciousness (i.e. inter-species experiential comparison) and simultaneously study the relationships between levels and contents of experience within the same species (i.e. intra-species experiential comparison). This will be done by changing the relation between two parameters in the model (i.e. $d<e$ or $d=e$).

This formal approach yields several implications for phenomenological studies (section \ref{discussion}), as well as discussions on tests and measurements of consciousness \cite{Bayne2024,Walter2021,Sandberg2010}, models of consciousness \cite{Signorelli2021b,Sattin2021,DelPin2021}, among others.

\section{Layers of Experience}\label{multi}

This section introduces the main assumptions within a multi-layer structure to model conscious experiences, followed by two compositional operations, $\otimes$ and $\odot$. With this mathematical structure, the model can scale from very simple postulates and eventually account for conscious experiences as combinations of a group of experiential aspects.


\subsection{Experiences as layers}

In the following, we consider to be conscious as simply \textit{experiencing}\footnote{Yet, consciousness require disambiguation, since different notions imply different types of theoretical and experimental questions: i) phenomenal consciousness refers to the \textit{what-it-is-like} to have a \textit{conscious} experience \cite{ThomasNagel1974}; ii) subjective experience is the first-person individual process of having an experience (\textit{what-it-is-like} sense) shaped by feelings, thoughts, and memories \cite{Gallagher2008}, sometimes also refereed as iii) lived experience, emphasising the personal and socio-cultural aspects of such experience \cite{Gallagher2008}; and finally, iv) content of consciousness as the "what" is experienced. In the following, we will refer to conscious experience in the first three senses, and make a distinction with the final sense.}, and model conscious experiences with a layered structure (Figure \ref{layersofexperience0}).

\begin{figure}[ht]
\centering
\tikzset{every picture/.style={line width=0.75pt}} 

\begin{tikzpicture}[x=0.75pt,y=0.75pt,yscale=-1,xscale=1]

\draw  [color={rgb, 255:red, 208; green, 2; blue, 27 }  ,draw opacity=1 ][fill={rgb, 255:red, 208; green, 2; blue, 27 }  ,fill opacity=1 ][line width=2.25]  (347.25,51) -- (422.5,51) -- (390.25,160) -- (315,160) -- cycle ;
\draw  [color={rgb, 255:red, 245; green, 166; blue, 35 }  ,draw opacity=1 ][fill={rgb, 255:red, 245; green, 166; blue, 35 }  ,fill opacity=1 ][line width=2.25]  (408.25,58) -- (483.5,58) -- (451.25,167) -- (376,167) -- cycle ;
\draw  [color={rgb, 255:red, 248; green, 231; blue, 28 }  ,draw opacity=1 ][fill={rgb, 255:red, 248; green, 231; blue, 28 }  ,fill opacity=1 ][line width=2.25]  (466.25,66) -- (541.5,66) -- (509.25,175) -- (434,175) -- cycle ;
\draw  [color={rgb, 255:red, 208; green, 2; blue, 27 }  ,draw opacity=1 ][fill={rgb, 255:red, 208; green, 2; blue, 27 }  ,fill opacity=1 ][line width=2.25]  (115.5,56) -- (142.5,56) -- (123,165) -- (96,165) -- cycle ;
\draw  [color={rgb, 255:red, 245; green, 166; blue, 35 }  ,draw opacity=1 ][fill={rgb, 255:red, 245; green, 166; blue, 35 }  ,fill opacity=1 ][line width=2.25]  (142.5,56) -- (169.5,56) -- (150,165) -- (123,165) -- cycle ;
\draw  [color={rgb, 255:red, 248; green, 231; blue, 28 }  ,draw opacity=1 ][fill={rgb, 255:red, 248; green, 231; blue, 28 }  ,fill opacity=1 ][line width=2.25]  (169.5,56) -- (196.5,56) -- (177,165) -- (150,165) -- cycle ;
\draw    (233.33,109) -- (264.33,109) ;
\draw [shift={(267.33,109)}, rotate = 180] [fill={rgb, 255:red, 0; green, 0; blue, 0 }  ][line width=0.08]  [draw opacity=0] (8.93,-4.29) -- (0,0) -- (8.93,4.29) -- cycle    ;
\draw [shift={(230.33,109)}, rotate = 0] [fill={rgb, 255:red, 0; green, 0; blue, 0 }  ][line width=0.08]  [draw opacity=0] (8.93,-4.29) -- (0,0) -- (8.93,4.29) -- cycle    ;

\draw (331,184) node [anchor=north west][inner sep=0.75pt]   [align=left] {\textbf{G}};
\draw (397,184) node [anchor=north west][inner sep=0.75pt]   [align=left] {\textbf{H}};
\draw (460,184) node [anchor=north west][inner sep=0.75pt]   [align=left] {\textbf{K}};
\draw (115,24) node [anchor=north west][inner sep=0.75pt]   [align=left] {\textbf{Experience}};

\end{tikzpicture}
\caption[Layers of experience.]{\small\textbf{Layers of experience.} Conscious experiences include several aspects, some of which are pre-reflective and implicit, and others are reflective and explicit. In the figure, an experiential layer (Experience) can include several of them, here represented by multiple colours (layers $G$, $H$, and $K$). Phenomenological investigations can disentangle them to understand the structural relation among aspects and experiences. Aspects of experience are also experiences, and therefore modelled by layers. Experiences with multiple aspects are multicoloured or multilayered, while aspects are monochromatic, representing a primitive/generator set of experiences.
}\label{layersofexperience0}
\end{figure}

Husserl's phenomenology recognised that conscious experiences can be investigated in levels or layers, from superficial to deeper ones \cite{Husserl1969,Husserl2001}\footnote{Other authors have also used layer metaphors, such as Freud's layers of the self (\cite{Freud1900}), MacLean's triune brain \cite{MacLean1990} and Signorelli's causal cognitive layers \cite{Signorelli2020a}. Theories of consciousness, such as Temporo-spatial Theory, also consider layered structures, for example, deeper spatiotemporal layers associated with the self \cite{Northoff2016,Northoff2017,Northoff2025}. Currently, the development of artificial neural networks, which also form a layer structure, continues inspiring layered models of experience.}. An "everyday" experience like tasting wine, for example, would be constituted by several implicit aspects (i.e. deeper layers), some of them including sensory modalities beyond taste (e.g. aromas, textures, colours). In Husserl's terminology, there are aspects of experience that are pre-reflective and others that are reflective \cite{Gallagher2008}. Pre-reflective aspects are implicit, not yet articulated for communication, but still constitutive of the experience. Reflective aspects, on the contrary, transform the pre-reflective implicit structure into communicable explicit "objects" of evaluation, allowing a system to assess its epistemic commitments and revise them accordingly \cite{Gallagher2008}.

In micro-phenomenological investigations \cite{Valenzuela-Moguillansk2019,Petitmengin2006,Petitmengin2018}, pre-reflective and reflective aspects are investigated through careful phenomenological interviews. This generates \textit{phenomenological data} that can be further analysed as an increasing chain of abstractions and categorisations, and mathematised as embedded layers (Figure \ref{layersofexperience0}). Chain of layers as aspects can be thought of as a set of primitives or generators of further experiences (i.e. multi-layers). In our general model, pre-reflective and reflective aspects of experiences are modelled as layers and multi-layers (see below).

For schematic purposes, one can represent aspects of experience with colours. For example, each colour can illustrate a sensory modality. In such an interpretation (other interpretations of aspects are also possible), each colour represents an experiential aspect that has an associated set of layers as specific instances of such an aspect (or sensory modality in our example below). For instance, orange colour can represent auditory experiences, and specific configurations account for distinguishable aspects like sound, harmony and melody. 

\includegraphics[scale=0.40]{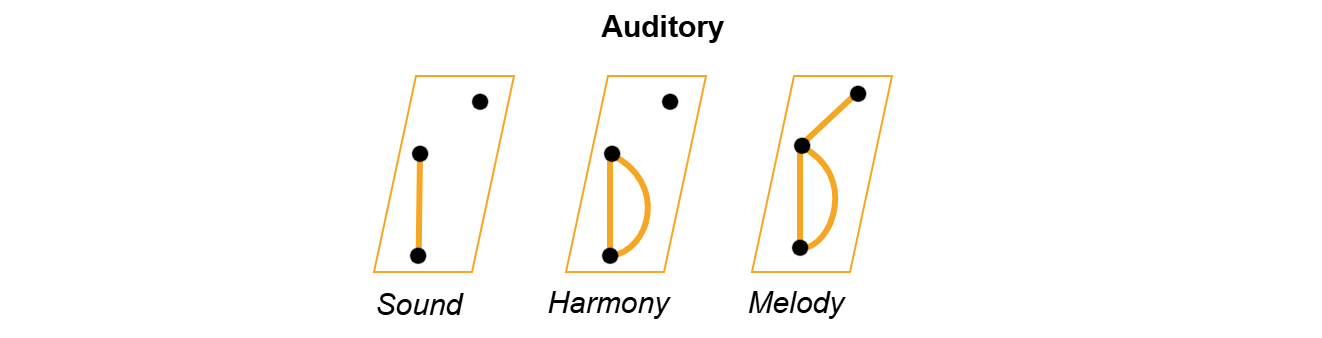}

We can also have other colours representing other aspects/modalities, such as red for taste. 

\includegraphics[scale=0.4]{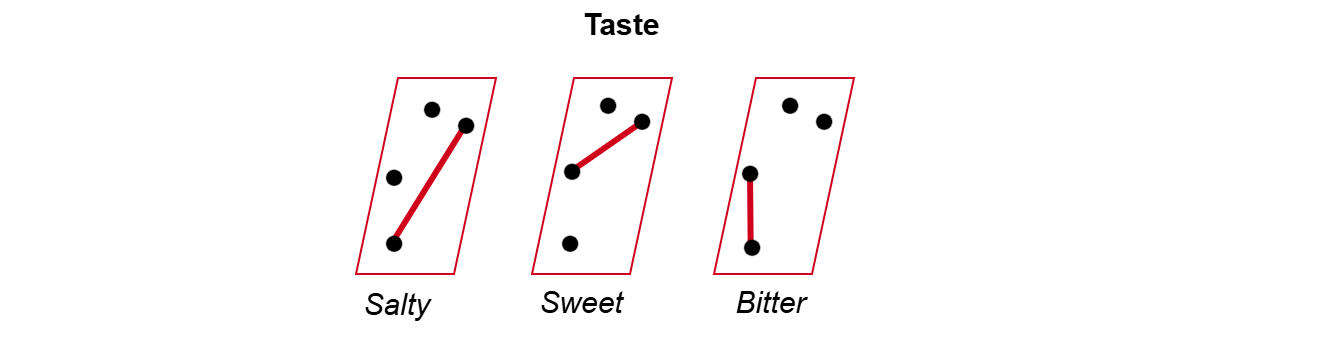}

Layers of experience include two mathematical structures: one is the phenomenological (the colours), and the other is the embodiment (the graph configurations representing biological or physical instantiations). For them, the relation can be expressed as:

\begin{prop}
The phenomenological structure indexes the embodiment structure. 
\end{prop}

This is coherent with the ideas expressed in \cite{Husserl1969,Husserl2001} where the phenomenological relations are previous to embodied one. Then, the colours form a set $C$ related to aspects of experience that indicate the aspects a certain graph contains. In other words, the structure of aspects appears before the structure of graphs. In fact, it should be seen as a \textit{labelling structure} to that of the graphs. That said, the phenomenological structure introduced here turns out to be an instance of what is formally defined in \cite{Kleiner2024} as a \textit{mathematical structure of conscious experience}. In our case, colours behave as an auxiliary structure to make a phenomenological sense of the dynamics of the graphs.

The definitive configurations of a group of experiences, colours/aspects and graphs, depend on informed and validated phenomenological data\footnote{The number and forms of these configurations depend on how much detail we add to the experiential description, i.e. the instantiations of layers of experience. This description is instantiated by the researcher \cite{Signorelli2024e}, following phenomenological principles and according to the particularities of the phenomenal structure to be modelled. This makes the role of the scientist explicit, as an active participant in the discovery scientific process \cite{Frank2024,DeJaegher2007}. Our structure works in analogy to the definition of a Coordinated system (e.g. Cartesian or Polar system). In short, the model requires being informed by phenomenological and rigorous considerations about what is expected to be modelled.}. For instance, validated phenomenological data could suggest that humans distinguish $j$ numbers of auditory experiences and/or their aspects (e.g., sound, harmony, melody) in a given context and $f$ numbers of tastes (e.g., bitter, sweet, salty). Then, for two aspects of experience and three instantiations each, we can use two colours (e.g., orange and red, as in the figures above) and define a set of layers for each aspect of the experience. 

Formally, we have a set $G_{orange}$ of layers $G_{orange}^{1}, G_{orange}^{2},...,G_{orange}^{j}$, with $j$ the total number of configurations for the aspect represented by the colour orange as audition in the examples above, and another set $G_{red}^{1}, ..., G_{red}^{f}$ for the aspect represented by red, e.g. taste.


These sets will be called \emph{layers of experience}. A layer corresponds to the pair $(G,col_G)$, where $G$ is a multigraph and $col_G$ is the mapping assigning colours/aspects to the edges in the multigraph $G$ (see \cite{Signorelli2024e} for a formal definition). For simplicity, we identify the layer $(G,col_G)$ with $G$ and, for $s\in\mathbb{N}$, 
we say that a layer is \emph{$s$-colored} if $col_G$ is onto and $s=|C|$. $s$ denotes the number of experiential aspects in the layer. In the example above, we have two aspects/colours $s=2$.

Each layer of experience is assumed as a closed unity, a single experience that is experienced as unified and may constitute further experiences. As such, we interpret layers as modelling a pre-reflective set of possible experiences and aspects \cite{Signorelli2024e}. These experiences can be experienced in potentiality or in actuality \cite{Husserl1960,Husserl1969}. 

\subsection{Experiences in potentiality}\label{expasmultilayer}

Building blocks of conscious experiences rarely appear isolated; some precede or follow others, some are embedded, and in general, they combine and compose to give rise to new configurations. In our model, experiential configurations are the result of the combination of layers as aspects of experience (which are also experiences). These configurations are depicted by merging layers of experience, such as:

\includegraphics[scale=0.40]{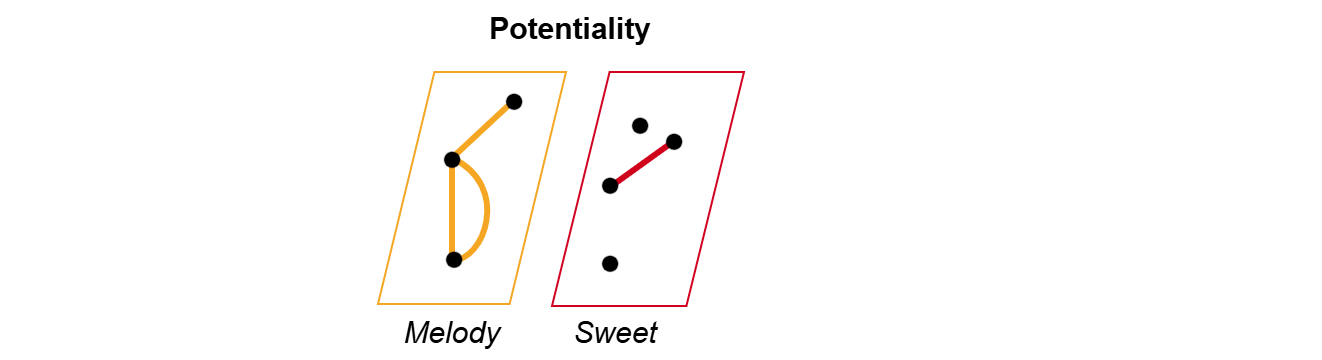}

Within Husserl's phenomenology, "Every subjective process has a process horizon, which changes with the alteration of the nexus of consciousness to which the process belongs ... an intentional horizon of reference to potentialities of consciousness that belong to the
process itself" (\cite{Husserl1960}, p. 44). These potentialities correspond to constitutive aspects of experience, and "we can ask any horizon what \textit{lies in it}, we can explicate or
unfold it, and \textit{uncover} the potentialities
of conscious life at a
particular time." (\cite{Husserl1960}, p. 45).

In our model, these potentialities are modelled as configurations of several layers. Formally, this requires the introduction of \textit{multilayers of experience}. We define the set of nodes indexed by the set $\{1,\dots,n\}$ and denote by $\mathcal{G}(n)$ the set of multigraphs with such $n$ nodes. Let $c$ be a single colour, then we denote by $\mathcal{G}^{c}(n)$ the set of $1$-coloured layers. Let $C=\{c_1,\dots,c_k\}$ be a set of colours/aspects, then we define the set of multilayer networks as the product\footnote{As prescribed in \cite{baez2020networkmodels} (Theorem 23, page 18) the different $\mathcal{G}$ are presented together in the form of a tensor product and, when instantiated as sets $\mathcal{G}(n)$ with their appropriate colour, we can use the usual Cartesian product. That structure is defined in \cite{baez2020networkmodels} in the language of Category Theory.}:

$$
\mathcal{G}^{\otimes C}(n):=(\mathcal{G}^{c_1}\otimes\cdots\otimes \mathcal{G}^{c_k})(n)
= \mathcal{G}^{c_1}(n)\times\cdots\times \mathcal{G}^{c_k}(n)
$$

Then, every multigraph in $\mathcal{G}^{\otimes C}(n)$ is called a \emph{$|C|$-coloured multilayer}. We can also have sets of multilayers in the form $\mathcal{G}^{\otimes C}(n) \times \mathcal{G}^{\otimes C'}(m)$, i.e. with different number of nodes (see examples in \cite{Signorelli2024e} for more details).

The tensor product $\otimes$ places layers in parallel and gives them a determined order, i.e. this product is non-commutative, meaning that $G\otimes H$ is not the same as $H\otimes G$. The implicit order given by $\otimes$ is a requirement associated with potentialities of experience and their interactions or combinations. These aspects have either the potential to interact or "as having a sense yet to be actualized" (\cite{Husserl1960}, p. 46). For example, "the perception has
horizons made up of other possibilities of perception, as
perceptions that we could have, if we actively directed the course of
perception otherwise..." (\cite{Husserl1960}, p. 46). The tensor also formalises the intuition of layers of experience $G$ and $H$ as \emph{experienced separated} or \emph{in parallel} and represents how the different experiential layers are presented without interacting between them (for more details see \cite{Sign,Signorelli2020c,Signorelli2020}). This could model, for example, specific elements or contents of our experiences that are pre-reflective and can be distinguished as "separated" aspects in a stream of experience. 

\subsection{Experiences in actuality}

According to Husserl's phenomenology, "every actuality involves its potentialities" (\cite{Husserl1960}, p. 44). In other words, "nothing exists for me otherwise than by virtue of the actual and potential performance of my own consciousness" (\cite{Husserl1969}, p. 234). Experiences are experienced right now and unified, i.e. in \textit{actuality}: "starting from the actually present intentionality
-the certainty, namely, that I could bring into play synthetically connected sequences of consciousness, with the unitary effect that I should continue to be conscious of the same object" (\cite{Husserl1969}, p. 234).

We model this by a commutative binary operation $\odot$ in $\mathcal{G}^{\otimes C}$ that combines layers or aspects of experience, unimodal or multi-layer, in a new single layer of experience \cite{Signorelli2024e}. The following example illustrates the composition operation $\odot$:

\includegraphics[scale=0.40]{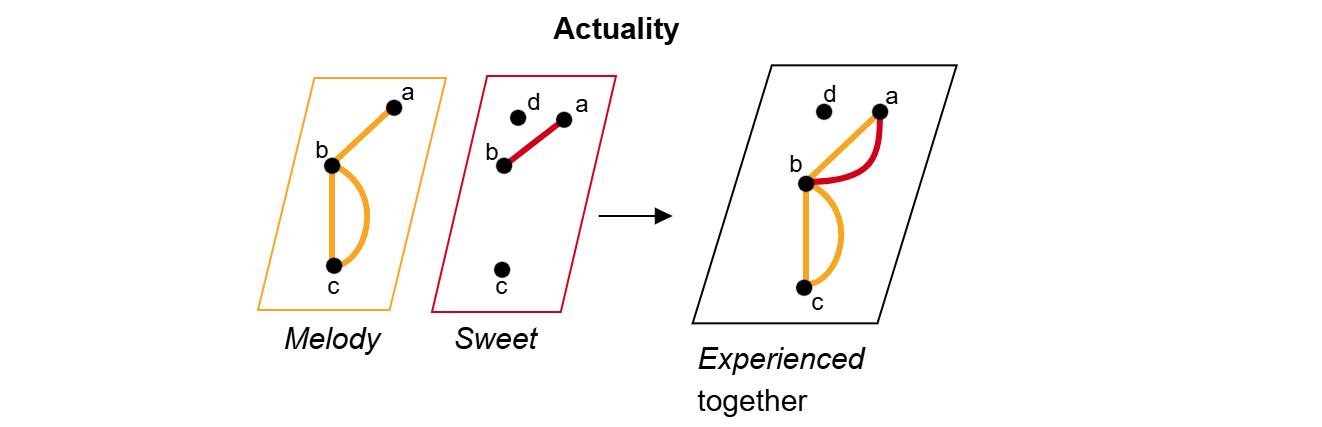}

Here we we label nodes for clarity. For other examples, check   \cite{Signorelli2024e}.

The operation $\odot$ corresponds to an accumulation of vertices and edges of two, or more, layers and multilayers of experience. This is also interpreted as \emph{experienced together} or forming a new unitary experiential content, as well as occurring in actuality, i.e. having two layers or multilayers of experience with the potential to interact; $\odot$ means that they are indeed interacting. The reader may also notice that this operation is commutative. 

The following is an application of the proposed model for a well-known example to show the concept of \textit{multistability} in Psychology.

\begin{example}
Let us formalise multistability in our layered model.
\vspace{1em}

\hspace{8em}
\includegraphics[scale=0.35]{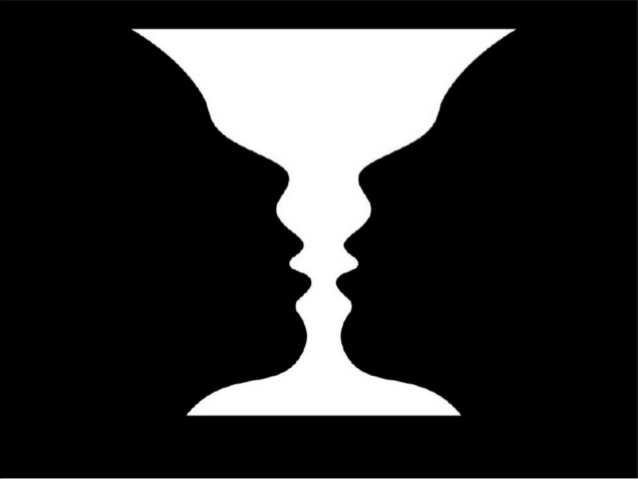}

\hspace{10em}The Rubin vase. Taken from \cite{Gestalt-LIV}.

\vspace{1em}
The model requires layers $L$ and $S$ related to modes of \textit{brightness} experience, and we assign the meanings: $L=light$, $S=shadow$. On the other hand, one may introduce an \textit{interpreter} layer $I=shape$ that mediates from the potentiality of $L$ and $S$ towards a certain visual interpretation. Putting all together, we obtain a three-layer chain $L\otimes I\otimes S$ for which:

\vspace{1em}

\tikzset{every picture/.style={line width=0.75pt}} 

\begin{tikzpicture}[x=0.75pt,y=0.75pt,yscale=-1,xscale=1]

\hspace{9em}
\draw    (339,193.77) -- (382.37,162.93) ;
\draw [shift={(384,161.77)}, rotate = 144.58] [color={rgb, 255:red, 0; green, 0; blue, 0 }  ][line width=0.75]    (10.93,-3.29) .. controls (6.95,-1.4) and (3.31,-0.3) .. (0,0) .. controls (3.31,0.3) and (6.95,1.4) .. (10.93,3.29)   ;
\draw    (321,195) -- (276.59,160.98) ;
\draw [shift={(275,159.77)}, rotate = 37.45] [color={rgb, 255:red, 0; green, 0; blue, 0 }  ][line width=0.75]    (10.93,-3.29) .. controls (6.95,-1.4) and (3.31,-0.3) .. (0,0) .. controls (3.31,0.3) and (6.95,1.4) .. (10.93,3.29)   ;
\draw  [fill={rgb, 255:red, 0; green, 0; blue, 0 }  ,fill opacity=1 ] (265,108) -- (268.25,108) -- (268.25,132) -- (274.75,132) -- (274.75,108) -- (278,108) -- (271.5,92) -- cycle ;
\draw  [fill={rgb, 255:red, 0; green, 0; blue, 0 }  ,fill opacity=1 ] (384,106) -- (387.25,106) -- (387.25,130) -- (393.75,130) -- (393.75,106) -- (397,106) -- (390.5,90) -- cycle ;

\draw (212,137.4) node [anchor=north west][inner sep=0.75pt]    {$\ \ \ \ L\odot I\otimes S\ \ \ \ \ \ \ \ \ \ \ \ L\otimes I\odot S\ \ $};
\draw (299,195.4) node [anchor=north west][inner sep=0.75pt]    {$L\otimes I\otimes S$};
\draw (255,67.4) node [anchor=north west][inner sep=0.75pt]    {$Cup$};
\draw (371,67.4) node [anchor=north west][inner sep=0.75pt]    {$Faces$};

\end{tikzpicture}

\vspace{1em}
\hspace{-1em}That is, for modes of brightness, the contribution of an interpreter is to shape them, as a result of the combination, allowing the emergence of a two-folded visual experience. In short, the model predicts that to model a bistable perception, one may require at least three phenomenological layers (and eventually three embodied ones). The exact nature of these layers requires phenomenological investigations, ideally using micro-phenomenological interviews. 
\end{example}

Summarising, our multilayer structure allows us to model the complexification of experiences via the composition of coloured layers. The structure contains phenomenological aspects represented by a set of colours and the embodied aspects, given by biological considerations. The main justification behind the selected operations is the formalisation of how aspects of experience are given in potentiality or actuality. While $\otimes$ shows unimodal experiences at our disposal in potentiality, $\odot$ speaks about a multimodal experience which is actually happening. It is expected that more $\otimes$ convert into $\odot$ through time. The way we mix multigraphs with $\odot$ allows us to identify these modalities without the cumbersome configurations that the usual tensor product of graphs would show. From now on, we work only with this general structure of operations, without specifying their graph or embodied structure.

\section{The partial ordered experiential structure}\label{pomonoid}

In the previous section, we introduced two ways of composing layers of experiences: $\otimes$ and $\odot$. These two operations form a mathematical structure that can model an increasing phenomenological structure based on a group of experiences informed by phenomenological data. As such, the structure can easily scale up and enable us to compare different aspects of experience in a theory-independent way. These products are generalized as concatenations of the form $G_1\oslash^{1}\cdots\oslash^{d-1}G_d$ with $\oslash^{i}\in\{\otimes,\odot\}$ 
for $\left|C\right|=k$ and $i=1,\dots,d-1$. Here, $k$ is the number of colours and $d$ is the number of layers. Note that some layers can have the same colour, implying $k \leq d$. 

According to our formal definitions in \cite{Signorelli2024e,Sign}, $\odot$ is a commutative operation while $\otimes$ is not. This distinction creates a priority of $\odot$ over $\otimes$, that is: 

\[
G\otimes H\odot K=G\otimes(H\odot K)
\]

We use brackets to represent priority. This asymmetry among products allows us to introduce the formal notion of a partial order of experiences. Importantly, this order is not a hierarchy between layers\footnote{A hierarchy may imply notions such as "supremacy" between what is "up" and what is "down". Our model is agnostic to such interpretations. The model only claims that the combination of layers can account for or describe the building up of more aspects and experiential relations.}. 

\subsection{Experiential comparison}\label{comparison1}

The comparison of experiences can be done either among global states of consciousness (e.g. deep sleep, dream sleep, awake, etc) or among experiences within a global state (e.g. taste of wine, listening to a melody). The notion of \textit{levels of consciousness} requires the existence of distinguishable global states of consciousness and eventually, the existence of a graded and quantifiable scale of conscious experience (i.e. there are higher and lower levels), allowing the comparison between states. The notion of \textit{content of consciousness} assumes combinations of experiential aspects to account for qualitative dimensions of experience within levels \cite{Tsuchiya2024}. These two theoretical distinctions help us to study the variability among subjective experiences within and across individuals in several conditions.

As commonly assumed in mainstream consciousness science, we start by assuming experiences are quantifiable and comparable \cite{Bayne2024,Bayne2016}, and see how far we can get with such an assumption. The taste of wine may require distinctions among several flavours and textures, while the taste of salt does not. In terms of supposed levels of conscious experience, it seems that minimal states of consciousness are lower than those of fully awake subjects. These two examples correspond to different cases of experiential comparison. The first corresponds to the comparison among contents of experience, while the second is about comparing levels of consciousness. 

Our model integrates both notions in the same mathematical structure. A partial order structure of a given chain of layers arises from the asymmetry already present in the two experiential operations $\odot$ and $\otimes$. For example, given two multi-layers  $G\otimes H\otimes K$ and $G\odot H\otimes K$ both $\in \mathcal{G}^{\otimes C}$, our mathematical structure imposes that $G\odot H\otimes K$ is over $G\otimes H\otimes K$, in the sense of being able to experience more aspects in a combined manner. In this case, the $\odot$ operation transfers its priority to any given chain of layers of experience. 

By convention, $G\otimes H\otimes K\le G\odot H\otimes K$, comparing the chains component-wise:

\begin{tikzpicture}
$
G\otimes H\otimes K\le G\odot H\otimes K
\draw[<->] (-4.1,-0.05)  .. controls (-4.065,-0.7) .. (-1.7,-0.05);
\node[] at (-3.065,-0.65) {\tiny {$\leq$}};
\draw[<->] (-0.5,0.2)  .. controls (1.065,1.7) .. (-3.2,0.3);
\node[] at (-1.2,1.25) {\tiny {$\leq$}};
$
\end{tikzpicture}

\hspace{-1em}Therefore, layers of experience $G\odot H$ are considered \textit{more} than $G\otimes H$, in the sense of being experience in actuality and unitary.

This assumption is partially justified by considering that experiencing in actuality (i.e. $\odot$) is phenomenologically more intense than experiencing in potentiality (i.e. $\otimes$). For example, the potential to experience two layers ready to interact, e.g. smell and taste, is not the same as the same two layers interacting and experiencing smell and taste altogether, in a given experience in this very moment. Note that we interpret interactions via $\odot$ as something being or happening, while $\otimes$ has the potential of happening. In other words, the $\odot$ has experiential \emph{actuality} and $\otimes$ experiential \emph{potentiality}.

We can extend and formalise the above intuitions: experiential comparison requires a partially ordered set, also called a \emph{poset}, as a set with a binary operation $\le$ which is reflexive, antisymmetric and transitive (see examples in \cite{Birk}). Then, the relation $\le$ in $\mathcal{G}^{\otimes C}$ impose an order on the concatenations of multigraphs. 

\begin{defn}\label{deforder}
Let be $\{i_1,\dots,i_d\}, \{j_1,\dots,j_e\} \subseteq \{1,\dots,k\}$ and $\pi$ a permutation of $\{1,\dots,k\}$ with $k=|C|$. For every $\oslash$ and $\ominus$ belonging to the set $\{\otimes,\odot\}$, we write
\[
G_{\pi(i_1)}\oslash^{1}\cdots\oslash^{d-1} G_{\pi(i_d)}\le G_{\pi(j_1)}\ominus^{1}\cdots\ominus^{e-1} G_{\pi(j_e)}
\]
if and only if $\{i_1,\dots,i_d\} \subseteq \{j_1,\dots,j_e\}$ and there is no $l$ such that $\oslash^{l}=\odot$
and $\ominus^{l}=\otimes$.

\end{defn}

As will be formalised in the proof of the following Theorem, the reason for having a posetal structure for phenomenological experiences (and not an \textit{equivalence relation} or the structure of a \textit{preset}, for instance) is that, leaving aside reflexivity and transitivity, a poset is endowed with the antisymmetric property. Antisymmetry is interpreted here in the following sense: if everything that intervenes in a particular experience $A$ also participates in another experience $B$ and, reciprocally, the whole experience $B$ \textit{is contained in} the experience $A$, then $A$ and $B$ are the very same (perhaps multimodal) experience.  

\begin{thm}
$(\mathcal{G}^{\otimes C},\leq)$ is a poset.
\end{thm}

\begin{proof}
A poset must satisfy reflexivity, antisymmetry and transitivity. Let's check this in our structure for $\leq$:
\begin{itemize}{
\item{it is reflexive by taking $d=e$ and every $i=j$ in Definition \ref{deforder},}
\item{it is antisymmetric since, whenever we have chains  \[
G_{\pi(i_1)}\oslash^{1}\cdots\oslash^{d-1} G_{\pi(i_d)}\le G_{\pi(j_1)}\ominus^{1}\cdots\ominus^{e-1} G_{\pi(j_e)}
\]
and
\[
G_{\pi(i_1)}\ominus^{1}\cdots\ominus^{d-1} G_{\pi(i_d)}\le G_{\pi(j_1)}\oslash^{1}\cdots\oslash^{e-1} G_{\pi(j_e)},
\]
then we can argue that $d=e$ and every $\oslash$ is the same as every respective $\ominus$ and}
\item{it is transitive by taking 
\[
G_{\pi(i_1)}\oslash^{1}\cdots\oslash^{d-1} G_{\pi(i_d)}\le G_{\pi(j_1)}\ominus^{1}\cdots\ominus^{e-1} G_{\pi(j_e)}
\]
and 
\[
G_{\pi(j_1)}\ominus^{1}\cdots\ominus^{e-1} G_{\pi(j_e)}\le G_{\pi(k_1)}\circledast^{1}\cdots\circledast^{f-1} G_{\pi(k_f)}
\]
for which $\{i_1,\dots,i_d\} \subseteq \{j_1,\dots,j_e\} \subseteq \{k_1,\dots,k_f\}$ and it is obvious that there is no $l$ such that $\oslash^{l}=\odot$
and $\circledast^{l}=\otimes$.
}}.
\end{itemize}   
\end{proof}

The previous definition allows us to compare different chains of experiential layers. However, this comparison is partial. 

\begin{example}\label{example_chains}
The partial character of this ordering is shown by observing that:
$$G\otimes H \otimes K \leq G\otimes H \otimes K \otimes L \leq G\odot H \otimes K \otimes L$$
and
$$G\otimes H \otimes K \leq G\odot H \otimes K \leq G\odot H \otimes K \otimes L$$
while
$$G\odot H \otimes K \not\leq G\otimes H \otimes K \otimes L$$ 
and
$$G\otimes H \otimes K \otimes L \not\leq G\odot H \otimes K$$

\vspace{1em}
\end{example}

In the most general case, the reader may notice that we are comparing chains of layers of experience with different numbers of layers and operations between them (see section \ref{evolution} for implications). In this case, it is not always possible to compare chains of layers in terms of \textit{bigger} or \textit{lower}, i.e. comparison is partial (see section \ref{nocomparison}). Therefore, our first formal result points to the partial nature of the comparison of layers of experience, i.e. this simple form of quantification already restricts the kind of experiences we can compare and how.

\begin{example}
Let's assume three layers for three colours experiences $S$,$M$ and $L$, as usually assumed in the human case: S-cones for short-wavelength (blue light), M-cones for medium-wavelength (green light), L-cones for long-wavelength (red/yellow light) and their respective phenomenologies. The model entails that experiencing $S \odot M \otimes L$ is not fully comparable with $S \otimes M \odot L$. Both are equally actual and unified, yet different experiences, phenomenologically speaking. By extension, experiencing colour relative similarity in one configuration is not more actual and unified than experiencing relative similarity in another experiential configuration.
\end{example}

Moreover, according to definition \ref{deforder}, we can interpret case $d=e$ as accounting for the order and experiential comparison within and among individuals of the same species (i.e. levels of consciousness and phenomenal content intra-species), while case $d<e$ points to the order among different species (i.e. animal consciousness inter-species). We can assume this considering that the same species may share similar aspects of experience (at least in potentiality) and, therefore, layers of experience, while different species may not. This could be informed by phenomenal data (e.g. animals that do not recognise some colours) or biological data (e.g. different brain morphology, etc). Then, specific instantiations of this general structure may provide an interesting account of comparisons intra-species (see section \ref{nocomparison}) and inter-species (see section \ref{evolution}).

\begin{example}\label{dogmonkey}
Let's imagine two mammalian species, dogs and monkeys, sharing most but not all experiential layers. Dogs may have $L_{1}\otimes L_{2} \otimes L_{3} ...\otimes L_{n}$ and monkeys additional $A$ and $B$ layers, i.e. $L_{1}\otimes... \otimes L_{n} \otimes A \otimes B$. If a dog is experiencing the maximal interacting state by having all the layers with $\odot$, we have: $L_{1}\odot L_{2} ...\odot L_{n}$. Such a structure is equal to $L_{1}\odot L_{2} ...\odot L_{n} \otimes A \otimes B$ in the monkey case (note that layers $A$ and $B$ are just in potentiality, while the rest are in actuality). In other words, more aspects in potentiality do not entail more experiences in actuality, since in certain cases and contexts, a global maximum of one species (e.g., a dog in this example) may be equal to or \textit{bigger} than a local maximum in another species (e.g. monkeys in this example).
\end{example}

\subsection{Experiential trajectories}\label{comparison2}

The most general case of experiential comparison (i.e. $d\leq e$, intra and inter-species simultaneously), yields very complicated diagrams. For simplicity, we will focus on $d=e$.

The partial order in definition \ref{deforder} can be extended to account for experiential trajectories, or flow of experience, by introducing mappings $f_j$ that take one operation $\otimes$ of index $j$ and transform it into a $\odot$ operation. This partial order allows us to define the following\footnote{To simplify 
the notation, we sometimes use lowercase letters $x$ and $y$ to denote multilayers of $\mathcal{G}^{\otimes C}$.}.

\begin{defn}
Let the mapping $f_{j}\colon \mathcal{G}^{\otimes C}\to \mathcal{G}^{\otimes C}$:
\[
f_{j}(x)=\begin{cases}
G_{1}\oslash^{1}\cdots G_{j}\odot G_{j+1}\cdots\oslash^{d-1}G_{d} & 
\textrm{if }x=G_{1}\oslash^{1}\cdots G_{j}\otimes G_{j+1}\cdots\oslash^{d-1}G_{d}\\
x & \textrm{otherwise}
\end{cases}
\]
for $j=1,\dots,d-1$. We say that 
$x,y\in \mathcal{G}^{\otimes C}$ are \emph{comparable through $f_j$} if $f_j(x)=y$.
\end{defn}

For example, $f_1$ and $f_2$ act such as:

$$\xymatrix{G\odot H \otimes K & & G\otimes H \odot K \\ & G\otimes H \otimes K \ar[ur]|{f_{2}} \ar[ul]|{f_{1}} &}$$

For clarity, we use the notation $f_{j}$ for any mapping defined above, avoiding the list of indexes and adding $f_0$ as the identity.

We extract two immediate consequences from how $f_{j}$ works. The first establishes that one can obtain the top element after an action of every $f_{j}$ over a given concatenation whatever the ordering may be:

\begin{prop}
$f_{i_{1}}\cdots f_{i_{d}}(G_{1}\oslash^{1}\cdots\oslash^{d-1}G_{d})=G_{1}\odot\cdots\odot G_{d}$
for $i_{1}<\cdots<i_{d}$ a permutation of $1,\dots,d$.
\end{prop}

The second consequence is that the application of $f_{j}$ always increases.

\begin{prop}\label{top}
$f_{j}(G_{1}\oslash^{1}\cdots\oslash^{d-1}G_{d})\geq G_{1}\oslash^{1}\cdots\oslash^{d-1}G_{d}$.
\end{prop}

These mappings $f_{j}$ are also order-preserving, i.e. if a multilayer interaction has fewer aspects in actuality than another, such order is preserved after applying any $f_{j}$ to both multilayers. For example, for $G\otimes H \otimes K \leq G\odot H \otimes K$ we have $$f_j(G\otimes H \otimes K) \leq f_j(G\odot H \otimes K)$$ 
for $j=1,2$.

With this mathematical structure, let's assume a group of experiences starts at a first-time point from which they are disentangled, in potentiality and parallel configuration, i.e. they operate using $\otimes$. Then, experiential layers can start to be combined via applications of $f_{j}$. This process generates an increasing flow from lower to upper configurations. The next example illustrates how $f$ works for longer chains and how they describe a \emph{flow} and potential trajectories on the poset of experiences $\mathcal{G}^{\otimes C}$.

\begin{example}\label{example3}
Given three multilayers $G$, $H$ and $K$ and all the possible ways of interacting using $\odot$ and $\otimes$, we obtained the next possible flow of experiences: 
{\tiny 
\[
\xymatrix{ &  & & G\odot H\odot K\\
G\odot H\otimes K\ar[urrr]|{f_{2}} & G\odot K\otimes H\ar[urr]|{f_{2}} & 
H\odot K\otimes G\ar[ur]|{f_{2}} & G\otimes H\odot K\ar[u]|{f_{1}} & 
H\otimes G\odot K\ar[ul]|{f_{1}} & K\otimes G\odot H\ar[ull]|{f_{1}}\\
G\otimes H\otimes K\ar[u]|{f_{1}}\ar@/_{4pc}/[urrr]|{f_{2}} & 
G\otimes K\otimes H\ar[u]|{f_{1}}\ar@/_{5pc}/[urr]|{f_{2}} & 
H\otimes G\otimes K\ar[ull]|{f_{1}}\ar@/_{3pc}/[urr]|(.24){f_{2}} & 
H\otimes K\otimes G\ar[ul]|(.32){f_{1}}\ar[ur]|{f_{2}} & 
K\otimes G\otimes H\ar@/^{4pc}/[ulll]|{f_{1}}\ar[ur]|{f_{2}} & 
K\otimes H\otimes G\ar@/^{4pc}/[ulll]|{f_{2}}\ar[u]|{f_{2}}
}
\]
}
\end{example}

As stated above, these trajectories formalise the intuition that experiences build up. Some experiences precede others, as primitives or generators of others. 

Example \ref{example3} also allows us to introduce a formal notion of levels of conscious experience. First, layers of experience interacting by $\odot$ or $\otimes$ are interpreted as instances of phenomenal content. Then, we can organise all horizontal layers of experience as a disjoint union of \emph{levels} according to the number of $\odot$ appearing in every concatenation. These horizontal configurations, in a poset, define levels of conscious experience, in which transformations $f_{j}$ account for sequential transitions, while $f_{ij}$ accounts for experiential "jumps". These experiential jumps could also be assembled as experiential groups, defining global states of consciousness incorporating several levels of experience, or sub-levels, within them. These sub-levels may be associated with experiences that share some common aspects, although in different configurations (e.g. the taste of wine and a melody, or the taste of cheese and the same melody).  Importantly, every two chains of layers living on the same level (horizontal lines in example \ref{example3}), although not comparable using $\leq$, contain the same number of appearances of $\odot$. This means that, according to definition \ref{deforder}, experiences at the same level cannot be compared in terms of more or less aspects. 

Now, one may also expect a mapping to go back and forth across experiences. In this case, we may consider an inverse transformation $g_j$. For example, $g_{1}$:

\[
g_{1}(G\odot H\otimes K)=G\otimes H\otimes K
\]
Yet, this transformation loses the well-definedness condition for the non-commutativity of $\otimes$, giving two possible results: 

$$\xymatrix{& G\odot H \otimes K=H\odot G \otimes K\ar[rd]|{g_{1}}\ar[ld]|{g_{1}}  &\\  G\otimes H \otimes K && H\otimes G \otimes K }$$

Here, $G\odot H \otimes K$ and $H\odot G \otimes K$ are considered equal in terms of aspects, actuality, and interactions, since i) $\odot$ is a commutative operation and ii) it has priority over $\otimes$. 
In other words, the two expressions have similar degrees in actuality and unified experiences, and pertain to the same level of experience. However, the result of applying $g_{1}$ gives us two results that are not equal any more (since $\otimes$ is not commutative). Therefore, the mappings $g_j$ are not well-defined functions, but relations. This mathematical "disadvantage" turns out to be an advantage for our model: being not well-defined mappings means that an inverse transformation has the potential to send an experience to several other configurations without knowing exactly what that would be, i.e. our mathematical structure can predict the next experiences if they build up (using $f_j$), but imposes a constraint, or undetermined condition if the experience goes back to minimal lower ones (using $g_j$). This behaviour of $g$ is another conjecture/prediction of the model that can be tested: given a set of noncommutative layers of experience and their phenomenal quality, an inverse operation after their composition might end up in different configurations than the original case. For example, “unbinding” after examples of “attentional binding” should result in a set of diverse configurations, potentially predicted by our model. As far as we are aware, experiments of this kind have not been performed yet.

Finally, $f_j$ and $g_j$ are interpreted as pre-reflective and reflective operations, respectively, such as the first build-up experiences, while the second disentangle experiential aspects. The reflective operations, as usually discussed in phenomenological literature and qualitative research, can reconfigure existing ones, and our inverse $g_j$ accounts for such reconfigurations.

In short, a poset structure constrains how experiential layers evolve among a chain of experiences (see section \ref{discussion} for interpretations and implications) and imposes restrictions to predictability conditions.
 
\begin{section}{Implications and discussion}\label{discussion}

In this section, we discuss the main results and their consequences. In short, structural and mathematical choices constrain the comparison of phenomenal experiences.

\subsection{Implications for experience comparison}\label{nocomparison}

One would like to know if experiencing the taste of wine entails more experiential aspects than seeing a beautiful landscape, or perhaps being able to distinguish between human experiences and animal experiences. In the first case, an important limitation is translating how one specific experience is \textit{more} actual and \textit{more} unified, than other experiences, e.g., by using verbal reports, neural measurements, or other techniques. In the second case, since animals can not provide verbal reports, experiential comparisons among species are challenging \cite{Leopold2003}. By extension, these difficulties also arise when assessing whether patients in minimal states of consciousness are minimally aware, and regarding pre-verbal children and babies.

\begin{example}\label{example_wine}
Let's assume that "tasting wine" is a subjective experience that requires at least one taste layer/aspect ($T_i$). Within such experience, wine may require two instantiations, bitter ($B$) and fruity ($F$), such that the experience, in potentiality, is given by $T \otimes B \otimes F \otimes ... T_{n}$. The same wine may trigger a specific configuration to a lay person, e.g. $T \odot B \otimes F \otimes ... T_{n}$, while a sommelier, as an expert, may experience a more elaborate one $T \odot B \odot F \odot ... T_{n}$. In this case, even though both experiences have the same number of layers/aspects, the first is \textit{less} actual than the second one. Phenomenologically, the first experience triggers fewer phenomenal distinctions than the second one.
\end{example}

According to section \ref{comparison1} and \ref{comparison2}, we note that under the assumption that experiences are comparable (i.e. the need to introduce a poset structure for such comparison) and described by layers of experience (i.e. a general structure to account for experiential configurations), we reached a counterintuitive result: an ordered structure to compare experiences leads us only to partial comparisons of inter-species experiences (section \ref{comparison1}) and intra-species experiences (section \ref{comparison2}). 

In other words, assuming: i) a poset structure, ii) the definition \ref{deforder} and iii) mappings $f_j$, then interactions given by the application of $f_j$ are the only ones comparable, i.e. between horizontal lines or levels in example \ref{example3}, but not between configurations within the same levels. This is because, qualitatively, configurations at the same level are considered different; however, simultaneously, there is no mathematical way, in this structure, to quantify the difference between, for example, $G\otimes H \otimes K$ and $G\otimes K \otimes H$. This mathematical constraint extends to other experiential configurations. For instance, the form $G\odot H \otimes K$ is less actual than $G\odot H \odot K$. Yet, $G\odot H \otimes K$ is not comparable with $G\otimes H \odot K$, since these configurations do not have a defined order within the structure (i.e. only configurations satisfying the conditions in definition \ref{deforder} are comparable). This structural and mathematical constraint makes it impossible to quantify and compare (bigger or smaller than other) layer configurations at the same level. 

\begin{example}
Let's consider "tasting wine" as a subjective experience $T \otimes B \otimes F \otimes ... T_{n}$ as in example \ref{example_wine}, and seeing a "beautiful landscape" requiring visual layers $V$, beauty $Y$, and several layers that constitute the landscape $L_{m}$ such that $V \otimes Y \otimes L_{1} \otimes L_{2} \otimes... L_{m}$. In this toy example, seeing the landscape might have more aspects than testing wine, as far as we phenomenologically find $n \le m$. Otherwise, there is no way to say which one is more actual than the other.
\end{example}

This structural conclusion makes phenomenological sense, i.e. how could we distinguish a configuration of experiences having more actuality and unity than another configuration having the same group of aspects but ordered differently, or having the same number of experiential aspects but of a different kind?

In summary, the main consequence of our assumptions, the poset of experiential layers and definition \ref{deforder} (exemplified by examples \ref{example_chains} and \ref{example3}), is that experiential comparison requires a distinction between interacting layers that allow comparison within them, i.e. different levels on the chain of possibilities, and interacting layers that do not, i.e. configurations at the same level on the chain of possibilities. In other words, levels of consciousness are comparable (by counting the appearance of $\odot$) while phenomenal contents at the same level are not. 

This mathematical conclusion imposes a space of possible experiences and allows flows and trajectories among them. Case by case, this structure can be stated using inputs from phenomenological and/or biological data to study further consequences of phenomenological investigations.

\subsection{Implications for evolutionary and animal consciousness}\label{evolution}

The number of experiential layers (i.e. potential aspects to be experienced) and their combinations account for experiential systems, allowing partial comparison (section \ref{comparison1} and \ref{comparison2}). Depending on the layer structure, we can compare some experiential configurations inter-species ($d<e$) and intra-species ($d=e$). The first case allows us to study animal consciousness from a mathematical perspective (section \ref{comparison1}), while the second investigates levels and phenomenal content within individuals of the same species (section \ref{comparison2} and \ref{nocomparison}). 

In the most general case ($d \leq e$), the layer structure assumes that different species have different numbers of experiential layers, i.e. having different potentialities for experiencing. One can consider, for example, biological layers such as cortical organisations, or more broadly, biological organisations which may also constrain phenomenal possibilities (e.g. visual layers, tactile layers, etc). A simple example may be animals with two types of photoreceptors, e.g. one rod and one cone, and animals with one rod and three cones, plus additional cortical structure. As such, a way to study and compare experiential potentialities is through the instantiation of layers as analogous to biological organisation \cite{Signorelli2024e}, with additional justified assumptions and inferences that can be mathematised as further structure in the model.

Then, our mathematical analysis suggests that being \textit{upper} on the \textit{evolutionary} scale does not always entail a \textit{bigger} consciousness (in the sense of more actual and unitary), as something graded on an absolute and unique scale. 

In that case, the experiential configuration is the one with more layers being experienced in actuality. As we see in the examples \ref{example_chains} and \ref{dogmonkey}, some cases cannot be compared, i.e. we cannot say which one is \textit{bigger} in an absolute sense. For instance, the experiential configuration $G\odot H \otimes K$, despite having a $\odot$ operation, is not \textit{bigger} than $G\otimes H \otimes K \otimes L$, which has more layers, in an absolute sense. In other words, $G\odot H \otimes K$ and $G\otimes H \otimes K \otimes L$ do not have an order relation: while $G\odot H \otimes K$ has more experiential \emph{actuality} than $G\otimes H \otimes K \otimes L$, the latter has more experiential \emph{potentiality} than the former. 

In the inter-species case, it means that even though a species may have the potential for several modes of experience, specific instantiations may not always reach its local maxima (local for the species, in comparison to all possible layer configurations defining a global maxima), and therefore not comparable to species that may have reached their local maxima. In other words, since layers describe and represent the possibility of experiences, there are local maxima and partial maxima (partial within the layers below the local maxima of a species) that species may reach independently. Each species may have its maxima, some of which are comparable and others not. Moreover, for example, experiencing relative similarity in one species can correspond, structurally, to a subset of layer configuration in another species (e.g., colour similarity), but it can also be the case that some layers are present only in the \textit{lower} species, but overall have fewer layers (e.g. odour experiences of rats compared to humans). This makes some aspects comparable but others not, and might be suggested, for instance, by extrapolations of behavioural experiments in rats \cite{Nakayama2022} and monkeys \cite{OpdeBeeck2001}. 

All this seems coherent with current biological evolution theories. Evolution is not a linear process, e.g. ants may be considered at the top of an evolutionary chain without having a straightforward comparison with mammals (at least as the world's most successful eusocial insects \cite{Ward2014,Johnson2013,Cai2024}). As such, assuming that humans or other animals are \textit{upper} in the chain of evolution and that this implies always a higher-graded conscious experience, seems to be a theological bias. Mathematically, and under the assumptions of our layer structure, our model entails different modes of phenomenal experience (i.e. different species might be conscious of different numbers of experiential aspects and their combination), but does not always entail, by itself, \textit{more} actuality and unitary character of such experiences.

By extension, animal experiences and perhaps even synthetic experiences might not always be comparable to human experiences (e.g. the "colour" experience of an octopus). 


\section{Conclusions}

We developed a simple layer model to study the structure of experiential comparisons, which can complement and extend cognitive and philosophical approaches by adding extra structure to the layers of experience (e.g. metrics). Our model has a minimal set of assumptions: experiences can be accounted for by layers (either phenomenological representations or biological embodied implementations), layers can be composed, and composition can be actual ($\odot$) or potential ($\otimes$). From these basic assumptions, the poset structure and definitions follow naturally under the assumption that experiences are comparable. The layer of experiences model, introduced through section \ref{multi} and \ref{pomonoid}, can also scale to address further assumptions of experiential comparison: the modeller only needs to add extra structure, either in the form of geometries, metrics, and others, to study the consequences of such extra assumptions.

Our model is originally inspired by the enactive and embodied approach \cite{Varela2016,Signorelli2024e}. Yet, also finds connections with developments such as \textit{Spatio-temporal Neuroscience} (see \cite{northoff2024}) and \textit{Dynamic Layer Model of the Brain} (\textit{DLB}). Our model is close to DLB in the sense that it makes use of layers that \textit{rotate} and interact to formalise the structure of conscious experience (see \cite{Signorelli2024e}). In that sense, the model resembles the idea of oscillation, also present in DLB. At the same time, the model is only partially related to DLB since, although our model conveys a structural formalisation valid for neuronal systems, the model is not restricted to neural systems, and, differently than other models, our approach considers an indexing structure that makes explicit the primary role of phenomenological aspects (see section \ref{multi}).

In this article, the consequences of our model and structural assumptions point to: i) a general enough structure to account for inter-species and intra-species experiential comparisons based on a few assumptions, ii) under these assumptions, experiences are partially comparable, not always comparable in absolute terms, iii) the evolution of conscious experience may entail different modes of phenomenal experiences, not always implying \textit{more} consciousness. Our characterisation also accounts for structural restrictions on experiential comparisons and experiential trajectories imposed by our means of modelling and mathematical choices. In this sense, mathematics has the power to instantiate assumptions uniquely and transparently and study their consequences. This makes our methodology a powerful tool that also acknowledges the importance of the scientist's experience, as an active modeller of the phenomenon intended to be modelled.

Finally, structural approaches, in general, are well-suited to capture categorical aspects of experience; yet, an important limitation is that one cannot rule out aspects of experience that may not lend themselves to such a categorical description \cite{Wachtel2022}. Other complementary models/approaches (e.g. arts) might describe those, if at all. Moreover, our model does not make ontological statements, and, on the contrary, it should be instantiated from concrete phenomenological descriptions that are not mathematical in nature (i.e. from phenomenological interviews or phenomenological investigations). As a limitation, or a virtue (depending on the perspective), the identification of primitive experiences, and the number/forms of fine-grained graph configurations depend on the participatory scientist acquiring such experiential descriptions. As such, by using a formal mathematical model, we are studying the consequences of a particular set of assumptions (i.e. our postulates). Yet, according to Husserl's phenomenology, any set of postulates (and axioms) is not inclusive enough to explain all aspects of conscious experience. A model only explains and predicts structural features in their specific instantiation. 

We end by remarking that using Category Theory (\cite{McLane}), posets become categories, and we can prove that $f_{j}$ are idempotent endofunctors corresponding to \emph{monads} and modal operators. In this future extension,  each new experience can be interpreted as a modality in a logic system, having a multimodal system where every modality $\diamondsuit$ is a conjunction of possible applications of functions $f_j$ satisfying a group of axioms. In future works, we exploit the particularities of modal logic applied to the analyses of structural aspects of conscious experience \cite{Diaz-Boils2025}.

\end{section}

\section*{Acknowledgment}

The authors thank the Association for Mathematical Consciousness Science (AMCS) and the Models of Consciousness Conference for gathering some of the discussions that motivated this research. CMS was supported by FNRS-Belgium through its grant Embodied - Time - 40011405 and Carlsberg Foundation grant number CF22-1432 during part of this research. NT was supported by the National Health Medical Research Council (APP1183280), Australian Research Council (DP240102680) and Japan Society for the Promotion of Science, Grant-in-Aid for Transformative Research Areas (23H04829, 23H04830).

\end{document}